\documentclass[aps,twocolumn,showpacs,preprintnumbers,floatfix]{revtex4}
\usepackage{graphicx}
\usepackage{epsfig}
\usepackage{amsmath}
\usepackage{dcolumn}
\usepackage{bm}
\newcommand{\pslash}{p \!\!\!/}
\begin{document}
\title{Oscillatory behavior of the domain wall fermions revisited}
\author{\small Jian Liang,$^{1}$ Ying Chen,$^{1,}$\footnote{cheny@ihep.ac.cn} Ming Gong,$^{2}$ Long-Cheng Gui,$^{1}$ Keh-Fei Liu,$^{2}$ \\
Zhaofeng Liu,$^{1}$ and Yi-Bo Yang$^{1}$}
\affiliation{\small $^1$Institute of High Energy Physics and Theoretical Center for Science Facilities, \\
Chinese Academy of Sciences, Beijing 100049, China\\
$^{2}$Department of Physics and Astronomy, University of Kentucky, Lexington, Kentucky 40506, USA}

\begin{abstract}

{ In the generic domain wall fermion formulation of chiral fermions on the lattice, the zero modes of the four-dimensional
Wilson fermion operator with the negative mass parameter $-M_5$ introduce unphysical massive modes propagating in
the four-dimensional spacetime. In the free fermion case, the pole mass of this kind of unphysical modes is given by $\tilde{E}=\ln(1-M_5)$, which
acquires an imaginary part, $i\pi$, when $M_5>1$ and results in an oscillatory behavior of the domain wall fermion propagator in time. The existence of the
unphysical modes in the presence of gauge fields is investigated in the mean field approximation, and their physical consequences are discussed.
In addition, we also give a semiquantitative criterion for tuning $M_5$ in the realistic numerical study.}
\end{abstract}
\maketitle

\section{INTRODUCTION}

The domain wall fermion (DWF), as a formulation for chiral fermions on the lattice, has been extensively
implemented in numerical lattice QCD studies near the physical pion mass. Although it has been successful in the realistic studies~\cite{RBC_review2010,RBC_review2012},
there is one question of the DWF which has not been clearly
addressed: oscillatory terms in the time dependence appear in hadron correlation functions
when DWFs are involved as either the sea or valence quarks~\cite{Dudek06,Walker-Loud,RBC_review2010}. Sometimes
this is attributed to either the artifact of the nonlocality of the DWF in four dimensions~\cite{Dudek06} or a cutoff effect~\cite{Walker-Loud}.
A study of the transfer matrix of DWF
on one space plus time and flavor dimensions~\cite{Negele} finds that, in the free case, negative eigenmodes arise when
the domain wall parameter $M_5$ takes the value $M_5>1$, which results in the oscillatory behavior of the quark propagator in time,
where the negative eigenmodes are obtained by solving numerically the eigenfunction of the DWF transfer matrix on a finite three-dimensional lattice.
However, it is not shown how this argument works for the five-dimensional case.
As far as the range of $M_5$ concerned in the lattice formulation of the DWF, it is known that chiral modes only exist for
$0<M_5<2$ in the free case. It is further argued that the optimal range of $M_5$ is $0<M_5<1$~\cite{Shamir:1993}.
In the case of the DWF coupled to gauge fields, the value of
$M_5$ can be shifted and should be tuned case by case for different types of the gauge action. This
tuning of $M_5$ was pioneered by the RBC group~\cite{Christ:1998} who paid attention to the quark
condensate $\langle \bar{\psi}\psi\rangle$ when varying $M_5$ and found a window of $1.65<M_5<2.15$
where $\langle \bar{\psi}\psi\rangle$ is nonzero and insensitive to $M_5$. A more sophisticated
method to tune $M_5$ is by using the spectral flow to obtain the lowest density of low eigenmodes of
the Hermitian Dirac operator $\gamma_5 D_W$ with $D_W$ as the Wilson fermion operator. The working
values of $M_5$ in the practical lattice QCD study range from 1.7 to 1.9~\cite{Aoki:2002,
Hagler:2007, Antonio:2007}.
\par
Intuitively, the oscillatory term should be related to the poles of the fermion propagator.  We study the
free DWF propagator in the momentum space and find that there exists another kind of singularity in addition to the pole corresponding to the chiral mode,
and then address the relation between this singularity and the oscillatory behavior of DWFs.
To address the more realistic case, we adopt the mean-field approximation by replacing the gauge links in the Dirac operator by their vacuum
expectation value, as has been done in Refs.~\cite{Aoki:1999,Feng06}.
We will also discuss the possible physical consequences of the unphysical modes in lattice simulations.

\par

This work is organized as follows: Sec.~\ref{section2} presents the study of the free fermion propagator
in the momentum space. The mean-field approximation of the DWF propagator in the gauge background is shown in Sec.~\ref{section3}.
Sec.~\ref{section4} contains the conclusion and discussion.

\section{DWF in the free case}\label{section2}
\subsection{Free DWFs in the continuum}
In order to investigate the spectrum of free DWFs, we start with Kaplan's original
proposal for the domain wall fermion operator in the continuum case~\cite{Kaplan:1992},
\begin{equation}\label{dirac5}
D_5=\gamma\cdot \nabla +\gamma_5 \partial_s + M(s),
\end{equation}
where $M(s)$ is a monotonic function on the fifth dimension $s$ with the asymptotic behavior,
\begin{equation}
M(s)=\left\{
\begin{array}{ll}
M  &  s\rightarrow \infty, \\
0  &  s=0, \\
-M &  s\rightarrow -\infty.
\end{array}
\right.
\end{equation}
Now we consider the zero modes $\Psi_0(x,s)$ satisfying $D_5\Psi_0=0$.
For a plane wave solution in the four-dimensional spacetime at $s\neq 0$, say,
$\Psi_0(x,s)=e^{ip\cdot x}u(p,s)$, in the chiral convention of $\gamma$-matrices,
\begin{equation}
\gamma_5=\left(
\begin{array}{lr}
I&0\\
0&-I \end{array}\right),~~ \gamma_4=\left(\begin{array}{lr}
0&I\\
I&0 \end{array}\right),~~ \gamma_i=\left(\begin{array}{lr}
0&-i\sigma_i\\
i\sigma_i&0 \end{array}\right),
\end{equation}
one can decompose the Dirac spinor $u(p,s)$ as
\begin{equation}
u(p,s)=\left(\begin{array}{l}\phi(p,s)\\\chi(p,s)\end{array}\right),
\end{equation}
where $\phi(p,s)$ and $\chi(p,s)$ are Pauli spinors, which subsequently satisfy the equations
\begin{eqnarray}
&&(ip_4+\sigma_i p_i)\chi+\frac{\partial}{\partial s}\phi+M\phi=0,\nonumber\\
&&(ip_4-\sigma_i p_i)\phi-\frac{\partial}{\partial s}\chi+M\chi=0,
\end{eqnarray}
or equivalently,
\begin{eqnarray}
&&p^2\chi-(ip_4-\sigma_i p_i)\left[\frac{\partial}{\partial s}+M\right]\phi=0,\nonumber\\
&&p^2\phi+(ip_4+\sigma_i p_i)\left[\frac{\partial}{\partial s}-M\right]\chi=0.
\end{eqnarray}
Obviously, for $p^2=0$, $\phi$ and $\chi$ decouple and only $\phi$ has normalizable solution for
$s>0$, which corresponds to the right-hand chiral mode bound on the domain wall. For $p^2\neq 0$, both
$\phi$ and $\chi$ satisfy the equation,
\begin{equation}\label{eigen}
\partial_s^2 (\phi,\chi)=(M^2+p^2)(\phi,\chi)\equiv \lambda^2 (\phi,\chi)
\end{equation}
with
\begin{equation}\label{lambda}
\lambda^2\equiv M^2+p^2.
\end{equation}
For $\lambda^2 \neq 0$, the general solutions take the form
\begin{equation}
\phi,\chi \sim e^{\pm \lambda s}.
\end{equation}
If $\lambda^2>0$, the normalizable solutions should be
\begin{equation}
\phi,\chi \sim e^{-|\lambda| s}
\end{equation}
with $|\lambda| = \sqrt{p^2+M^2}$, which damp exponentially along the $s$ dimension.
However for $p^2+M^2\le 0$, the normalizable solutions correspond to the scattering states
\begin{equation}
\phi,\chi \sim e^{\pm i|\lambda|s}.
\end{equation}

For the special case $p^2+M^2=0$, the solution corresponds to a constant mode in the $s$-dimension. In this case, the
four-dimensional dispersion relation is $-p^2=M^2$ which indicates a massive propagating mode along the
four-dimensional spacetime.

\subsection{Free DWF propagator on the lattice}\label{section2b}
For DWFs on a lattice, we start with the free DWF propagator in Shamir's formulation~\cite{Shamir:1993}.
Similar to the four-dimensional case, the lattice discretization of the five-dimensional Dirac operator $D_5$ in Eq.~(\ref{dirac5})
takes the Wilson's prescription by introducing the Wilson term with the Wilson parameter set to 1 to circumvent the fermion
doubling problem.

For the finite $s$-dimension with extension $L_s$, a domain wall and an antidomain wall is put at $s=0$ and $s=L_s-1$, respectively, with
a domain wall parameter $M_5$, such that the the right(left)-hand chiral fermions are bound on the spacetime slice at $s=0(L_s-1)$.
The bare current quark mass $m$ acts as the coupling of the right-hand chiral fermion at $s=0$ and the left-handed chiral fermion at $s=L_s-1$.
As such, the free domain wall fermion matrix in the four-dimensional momentum space (corresponding to the Euclidean spacetime and labeled by $\mu$) and $s$ is written as
\begin{eqnarray}\label{dirac_operator}
\hat{D}_{ss'}^{\rm{(m)}}(p)&\equiv& \theta(s)\theta(s')\theta(L_s-1-s)\theta(L_s-1-s')
D_{ss'}^{\rm{(0)}}(p)\nonumber\\ &+&
m\left[P_L\delta_{s,0}\delta_{s',L_s-1}+P_R\delta_{s',0}\delta_{s,L_s-1}\right],
\end{eqnarray}
where $D_{ss'}^{\rm{(0)}}(p)$ is the massless Dirac operator on an infinite fifth dimension, $s$:
\begin{equation}
D_{ss'}^{\rm{(0)}}(p)=\left[P_R\delta_{s+1,s'}+P_L\delta_{s-1,s'}\right]-\left[b(p)+i\tilde{\pslash}\right]\delta_{ss'}.
\end{equation}
In the above and the following equations we use the notations
\begin{eqnarray}\label{definition}
&&\tilde{\pslash}\equiv\sum\limits_\mu\gamma_\mu \sin p_\mu,\nonumber\\
&&\tilde{p}^2=\sum\limits_\mu \sin^2 p_\mu,\nonumber\\
&&P_{R/L}\equiv(1\pm\gamma_5)/2,\nonumber\\
&&b(p)=1-M_5+\sum\limits_{\mu}(1-\cos p_\mu),
\end{eqnarray}
where $M_5 = M a_s$ is the domain wall parameter. It should be noted that, for simplicity,
the expressions throughout
the article are in lattice units, taking the lattice spacings $a_s=a_\mu=1$. The function $b(p)$ comes from the hopping term
in the DWF action and depends directly on the domain wall parameter $M_5$, which takes the value in
the range $0<M_5<2$ for chiral modes to exist. In order to investigate the spectrum of DWFs in
Euclidean space, one usually introduces the second order operator $\Omega^{(0)}(p)$,
\begin{eqnarray}
\Omega_{ss'}^{\rm{(0)}}(p)&\equiv& \left[D^{\rm{(0)}}(p)D^{\rm{(0)}\dagger}(p)\right]_{ss'}\nonumber\\
&=&(1+b^2(p)+\tilde{p}^2)\delta_{ss'}-b(p)\left(\delta_{s,s'-1}+\delta_{s,s'+1}\right),\nonumber\\
\end{eqnarray}
which is Hermitian and non-negative. It is easy to check that, if
\begin{equation}\label{condition}
1+b^2(p)+\tilde{p}^2 \neq 2b(p),
\end{equation}
the two homogeneous solutions of $\Omega^{(0)}(p)$ are given as $e^{\pm \alpha(p) (s-s')}$ with $\alpha(p)$
defined as
\begin{equation}
\cosh \alpha(p)= \frac{1+b^2(p)+\tilde{p}^2}{2b(p)},
\end{equation}
which is actually the lattice version of Eq.~(\ref{lambda}) under the condition of
Eq.~(\ref{condition}) with $\alpha$ being the lattice counterpart of $\lambda$.
The inverse $G^{\rm{(0)}}(p)$ of $\Omega^{\rm{(0)}}(p)$ is thereafter expressed as
\begin{equation}
G^{\rm{(0)}}_{ss'}(p)=\frac{1}{2b(p)\sinh\alpha(p)}e^{-\alpha(p)|s-s'|}.
\end{equation}
Note that these discussions are based on the condition of Eq.~(\ref{condition}) which implies
$\sinh\alpha(p)\neq 0$.

With these prescriptions and the boundary conditions imposed in Eq.~(\ref{dirac_operator}),
the five-dimensional free quark propagator $\tilde{S}_F(p)$ can be
derived explicitly (the concrete procedure and the expressions, which can be found in
Refs.~\cite{Shamir:1993, Feng06}, are irrelevant to this work and therefore are omitted here).

The poles of $\tilde{S}_F(p)$ are given by the zeros of the function~\cite{Feng06}
\begin{eqnarray}
\Delta&\equiv&e^{2\alpha}\left(b-e^{-\alpha}\right)(1+m^2m_r^2)+(m^2+m_r^2)(e^{\alpha}-b)\nonumber\\
&+&2mm_r b(e^{2\alpha}-1),
\end{eqnarray}
and $m_r\equiv\exp(-\alpha L_s)$ is the so-called residual mass which accounts for the explicit
breaking of the chiral symmetry owing to the finite extension of the flavor dimension $L_s$. Letting
$\Delta=0$ one obtains the pole equation,
\begin{equation}\label{dispersion}
-\tilde{p}^2=W\left[(1+b^2+\tilde{p}^2)^2-4b^2\right],
\end{equation}
where $W$ is defined by $W=m_1m_3/m_2^2$ with
\begin{eqnarray}
m_1&=&(1+mm_r)^2,\nonumber\\
m_2&=&(1-m^2)(1-m_r^2),\nonumber\\
m_3&=&(m+m_r)^2.
\end{eqnarray}
Generally speaking, both $b$ and $m_r$ are functions of momentum, such that the dispersion relation
is very complicated. However, for a large enough $L_s$ and resultantly small enough $m_r$, in the
limit of $p^2\rightarrow 0$ and $m\rightarrow 0$ (in the units of inverse lattice spacing),
Eq.~(\ref{dispersion}) can be simplified as
\begin{equation}\label{pole}
-p^2=(m+m_r)^2(1-b^2)^2.
\end{equation}
which can be compared with the dispersion relation in the continuum Minkowski spacetime, and gives
the pole mass $M_P$ of the domain wall fermion
\begin{equation}
M_P\approx (m+m_r)(1-b^2).
\end{equation}
The physical quark propagator $S_F(p)$ in the four-dimensional Euclidean spacetime is realized from
$\tilde{S}_F(p)$ as

\begin{eqnarray}\label{regular}
  S_F(p)&=&P_L~\tilde{S}_F(p)_{0,L_s-1}P_L + P_L~\tilde{S}_F(p)_{0,0}P_R \nonumber\\
  &+& P_R \tilde{S}_F(p)_{L_s-1,L_s-1}P_L + P_R~\tilde{S}_F(p)_{L_s-1,0}P_R,\nonumber\\
\end{eqnarray}
which inherits the total divergence of $\tilde{S}_F(p)$. Therefore, when $m=0$ and $L_s\rightarrow
\infty$, this pole given by Eq.~(\ref{pole}) corresponds to a massless
chiral fermion in the physical spacetime.
The above results are well known in the literature. However, before we end the discussion of
this part, we would like to emphasize that the results above are all based on the condition
$1+b^2(p)+\tilde{p}^2\neq 2b(p)$ and there are no additional constraints on the value of $M_5$ apart
from $0<M_5<2$. What follows is the discussion of the consequence in the case of
$1+b^2(p)+\tilde{p}^2=2b(p)$.

\subsection{Unphysical propagating modes of the free DWF}
Superficially it appears that $1+b^2(p)+\tilde{p}^2> 2b(p)$ if $M_5\neq 1$, since the four
momentum takes values in the first Brilloun zone $-\pi/L_\mu\le p_{\mu}<\pi/L_\mu$ where $L_\mu$ is the
lattice extension in the $\mu$ dimension. This assumption is actually taken by default in the
previous works~\cite{Kaplan:1992, Shamir:1993}. However, the quark propagator in the coordinate
space is connected with that in the momentum space through a Fourier transformation; as such,
the integral over the momentum should be extended over the complex momentum space. In this situation,
there can be some complex momenta that satisfy the condition $1+b^2(p)+\tilde{p}^2= 2b(p)$.
Taking the static case ($p_i=0$) for instance, by combining with the definitions in
Eq.~(\ref{definition}), the solution of the equation
\begin{equation}
1+b^2(p_4)+\sin^2 p_4=2b(p_4)
\end{equation}
is given as
\begin{equation}
e^{\pm ip_4}=1-M_5.
\end{equation}
So the equality $1+b^2(p)+\tilde{p}^2= 2b(p)$ is permitted in the theory. Actually the zero
modes of the four-dimensional Wilson fermion operator $D_W(-M_5)$ with a negative mass parameter $-M_5$ satisfy the condition. In the momentum space, these zero modes $\psi_0(p)$ are given
by
\begin{equation}
D_W(p,-M_5)\psi_0(p)\equiv (b(p)+i\tilde{\pslash}-1)\psi_0(p)=0,
\end{equation}
with $p$ satisfying the above condition. Consequently a spinor $\Psi_s^{(0)}(p)=C \psi_0(p)$ satisfies the equation
\begin{equation}
D_{ss'}^{(0)}(p)\Psi_{s'}^{(0)}(p)=0,
\end{equation}
and gives a constant propagating mode along the $s$-dimension which is normalized as
$C=1/\sqrt{L_s}$ for a finite $L_s$. In other words, $\Psi_s^{(0)}(p)$ is a zero mode of $D_{ss'}^{(0)}(p)$, such that
there is no inversion for it and the deduction in Sec.~\ref{section2b} does not apply in this case. With respect to this point, the full
four-dimensional propagator of the DWF should take this kind of singularity into account in addition to the regular part $S_F(p)$
defined in Eq.~(\ref{regular}). Therefore, the free fermion propagator in the physical Euclidean space can be formally written as
\begin{equation}
\hat{S}_F(p)=\left\{%
\begin{array}{ll}
    S_F(p), & \left(1+b^2(p)+\tilde{p}^2\neq 2b(p)\right), \\
    S_F'(p), &\left(1+b^2(p)+\tilde{p}^2= 2b(p)\right), \\
\end{array}%
\right.
\end{equation}
where $S_F'(p)$ comes from the zero mode of $D_{ss'}(p)$ mentioned above.
The behavior of this fermion propagator with respect to the Euclidean time $t$ can be investigated
by performing the Fourier transformation over $p_4$. The propagator in the $t$ direction $\hat{S}_F(t)$
is obtained by the integration over the first Brillouin zone,
\begin{eqnarray}
\hat{S}_F(t)&=&\int_{-\pi}^{\pi}\frac{dp_4}{2\pi}e^{ip_4 t}\hat{S}_F(p),
\end{eqnarray}
which has contributions from both the physical poles described by Eq.~(\ref{dispersion}) and
the singularities due to the relation $1+b^2(p)+\tilde{p}^2= 2b(p)$, and the latter takes the form
\begin{equation}
{S}'_F(t)\sim e^{-\tilde{E}(p_i)t},
\end{equation}
where $\tilde{E}(p_i)\equiv-ip_4$ is the solution of the above relation. For $p_i=0$, one has
\begin{equation}\label{additional}
\tilde{E}(0)=-ip_4 =\pm \left[ \ln |1-M_5|+i\arg (1-M_5)\right].
\end{equation}
This corresponds to an additional propagating mode along the Euclidean time direction
apart from the physical mode discussed above. This is exactly the lattice version of the constant mode we discussed in
the continuum case. Because $M_5$ is defined as $M_5=Ma_5$ ($a_5$ here refers to the lattice
spacing in the fifth direction) the energy of this mode
\begin{equation}
\hat{E}(0)\equiv\ln|1-Ma_5|=Ma_5+O(a_5)~~~(a_5\rightarrow 0).
\end{equation}
Note that the constant mode and the scattering states discussed in Sec.IIA are normalized by $O(1/\sqrt{L_s})$, so for the
infinite $s$-dimension, say, $L_s\rightarrow \infty$, their contribution can be neglected. But for a finite $L_s$,
they have a sizable effect on fermion propagators. Since $M_5$ is a tunable parameter on the lattice, one can choose an optimal
$M_5$ which lifts this additional mode high enough in mass to decouple from the physical modes. Equation~(\ref{additional}) indicates
that $M_5\sim1$ meets the requirement in the free case. On the other hand,
when $M_5$ takes a value of $M_5>1$, the energy $\tilde{E}(0)$ given Eq.~(\ref{additional}) is complex, and the Fourier transform
will pick up the poles at
\begin{equation}
\tilde{E}(0)=\pm\left[\ln(M_5-1)+i\pi\right],
\end{equation}
which means this mode propagates in the time direction as
\begin{equation}
S(t)\propto (-1)^t \left(e^{-\hat{E}t}+e^{-\hat{E}(T-t)}\right),
\end{equation}
where $\hat{E}\equiv|\ln(M_5-1)|$. This is surely the origin of the oscillating mode in the temporal direction of hadron correlators. Given
the values of $M_5=1.1, 1.3, 1.5, 1.7$ as done in Ref.~\cite{Negele}, $\hat{E}$ takes the values
$2.303, 1.204, 0.693$ and $0.357$ respectively, which are exactly the energies of the negative
eigenmodes illustrated in Fig.~1 of that reference where a $2+1$-dimensional DWF transfer matrix
is numerically calculated. In fact, our discussion is for general spacetime
dimensions.
The discussion above also applies to the Borici's realization of the DWF~\cite{Borici:2000}. The only difference in this case is
that the massless domain wall fermion operator $D_{ss'}^{(0)}(p)$ in Eq.~(\ref{dirac_operator}) is
modified as
\begin{eqnarray}
D_{ss'}^{(0)}(p)&=&\left(2-b(p)-i\tilde{\pslash}\right)\left[P_R\delta_{s+1,s'}+P_L\delta_{s-1,s'}\right]\nonumber\\
&-&\left[b(p)+i\tilde{\pslash}\right]\delta_{ss'},
\end{eqnarray}
and the corresponding second order operator $\Omega^{(0)}(p)$ takes the form
\begin{eqnarray}
\Omega_{ss'}^{(0)}(p)&\equiv& \left[D^{(0)}(p)D^{(0)\dagger}(p)\right]_{ss'}\nonumber\\
&=&((b(p)-2)^2+\tilde{p}^2+(b(p)^2+p^2))\delta_{ss'}\nonumber\\
&-&(2b(p)-b(p)^2-\tilde{p}^2)\left(\delta_{s,s'-1}+\delta_{s,s'+1}\right).
\end{eqnarray}
Similar to Shamir's formalism, one can introduce a parameter $\alpha(p)$ defined by
\begin{equation}
\cosh\alpha(p)\equiv \frac{(b(p)-2)^2+\tilde{p}^2+b(p)^2+\tilde{p}^2}{2(2b(p)-b(p)^2-\tilde{p}^2)}.
\end{equation}
An unphysical mode exists in this case when
\begin{equation}
(b(p)-2)^2+\tilde{p}^2+b(p)^2+\tilde{p}^2= 2(2b(p)-b(p)^2-\tilde{p}^2),
\end{equation}
which is equivalent to
\begin{equation}
1+b(p)^2+\tilde{p}^2 = 2b(p).
\end{equation}
In other words, the unphysical modes also exist for Bori\c{c}i's domain wall fermions.
If one takes a closer look at the relation $1+b^2(p)+\tilde{p}^2= 2b(p)$, one finds it is nothing
but the equation for the zeros of the Wilson operator $D_W(-M_5)$ with the negative mass factor $-M_5$.
This means that the appearance of the unphysical modes is the direct consequence of the zero modes of $D_W(-M_5)$. There are also other formulations of DWF, such as the Optimal domain wall fermions~\cite{Chiu:2003, Chiu:2012} and the Mobius domain wall fermions~\cite{Mobius} which we will not address in the present work.

\subsection{The domain wall parameter $M_5$}
From the discussion above, one can see that the propagating behavior of domain wall fermions has
a close relation to the domain wall parameter $M_5$, which can be summarized as follows: i) The $\alpha$ depends on $M_5$. ii) The constant mode in the fifth dimension corresponds to an
unphysical mode in the four spacetime with energies dependent on $M_5$. To address the latter
it is desired that $M_5$ be close to 1, such that the unphysical mode is lifted very high and
decouples from physical modes. Note that there is no other requirement from this point.
As far as the first point is concerned, it demands that $\alpha$ should be as large as possible for all of
the fermion momentum in the four spacetime, such that the fermion modes bound on the two domains
have as little overlap as possible. This requires $b(p)$ approaching 0 for any $p$. Since
\begin{eqnarray}
b(p)&=&1-M_5 + \sum\limits_{\mu}(1-\cos p_{\mu})\nonumber\\
    &=&1-M_5 + 2\sum\limits_\mu \sin^2 \frac{p_\mu}{2},
\end{eqnarray}
$b(p)\rightarrow 0$ requires
\begin{equation}
M_5\rightarrow 1+2\sum\limits_\mu \sin^2 \frac{p_\mu}{2}>1
\end{equation}
for any $p$. For the momentum near the left corner of the first Brillouin zone, say, $p_\mu\sim 0$,
the optimal $M_5$ should be slightly larger than 1. Combining the above two constraints, the
reasonable range is $M_5\gtrsim 1$. In Ref.~\cite{Shamir:1993}, the author claims that $M_5$ should
be taken in the range $0<M_5<1$ in order to avoid the singularities for $b(p)=0$ in the
definition of $\alpha$. In retrospect, $b(p)=0$ is actually the ideal case, because the second order operator $\Omega^{(0)}$
is now a unit matrix (up to a factor) in this case whose inverse, $G^{(0)}_{ss'}\propto \delta_{ss'}$, shows that the
corresponding fermion modes are completely bound on the domain. In other words, it is favorable to have the singularities instead of unfavorable.
On the other hand, for $p^2\sim 0$, $M_5>1$
means there is a possibility that $b(p)<0$, such that $\alpha(p)$ can be complex,
\begin{equation}
\alpha(p)=|\alpha(p)|+i\pi,
\end{equation}
which means the propagating of fermion modes can be oscillatory in the $s$-dimension. However, if
the size of the $s$-dimension is taken to be $L_s={\rm even}$, the definition of the residual mass does
not change. So $M_5>1$ does not introduce physical problems for domain wall fermions.

\section{DWFs coupled to gauge fields}\label{section3}
Domain wall fermions are now applied extensively in the lattice QCD simulations. So in this section,
we turn to discuss the possible effects of the unphyiscal modes of DWFs in the presence of gauge
fields. In this case, the coupling term of fermion fields and gauge fields in the Shamir type
domain wall fermion action is written as
\begin{eqnarray}
S_{\rm int}&=&\frac{1}{2}\sum\limits_{x,s,\mu}\bar{\psi}_{x,s}\left[(1+\gamma_\mu)
U_\mu(x)\psi_{x+\hat{\mu},s}\right.\nonumber\\
&& +\left. (1-\gamma_\mu)U_\mu^\dagger(x-\hat{\mu})\psi_{x-\hat{\mu},s}\right],
\end{eqnarray}
where $U_\mu(x)$ is the gauge link at the spacetime site $x$ and in the spacetime direction $\mu$.
Obviously, $U_\mu(x)$ is also a function of spacetime coordinates and the interaction term will
make the concrete expression of the full propagator of the DWF in the momentum space more complicated.
In order to avoid the complication and maintain a semiquantitative discussion, we adopt the mean-field
approximation employed by Refs.~\cite{Aoki:1999, Feng06}, where the gauge links are replaced by
their vacuum expectation value $u_0$, i.e., the tadpole parameter~\cite{Lepage}. In the analytic
studies, the tadpole parameter $u_0$ is commonly defined by the vacuum expectation value of $U_\mu$
in a fixed gauge, for example, the Landau gauge: $u_0=\langle 1/3 {\rm ReTr} U_\mu\rangle_{\rm
Landau}$. In the numerical study, $u_0$ is always defined as $u_0=<1/3 {\rm ReTr} W_p>^{1/4}$ with
$W_p$ as the plaquette operator, and can be obtained in a self-consistent way in the Monte Carlo simulation.
However the $u_0$
is defined, it is found in the practical study that $u_0$ takes a value in the range $0.8\sim 0.9$
for most gauge actions in practice.  With this prescription, the discussions in the last section are also
valid but with the modification
\begin{equation}
\sin p_\mu \rightarrow u_0\sin p_\mu
\end{equation}
and
\begin{equation}
b(p)=1-M_5+\sum\limits_\mu (1-u_0 \cos p_\mu),
\end{equation}
such that
\begin{equation}
\cosh \alpha(p)= \frac{1+b^2+u_0^2\sum\limits_\mu \sin^2 p_\mu}{2b},
\end{equation}
while other expressions do not change. In this simplified case, the unphysical mode with $p_i=0$ is
given by $\sinh \alpha(p) =0$ with the energy
\begin{eqnarray}
\tilde{E}&=&-ip_4\nonumber\\
&=&\pm\left[\ln\left|\frac{4-M_5-3u_0}{u_0}\right|+i\arg\left(4-M_5-3u_0\right)\right]\nonumber\\
&\equiv& \pm \left[\hat{E}+i\arg\left(4-M_5-3u_0\right)\right]
\end{eqnarray}
which depends only on $M_5$ and $u_0$, where $\hat{E}=\ln\left|{(4-M_5-3u_0)}/{u_0}\right|$. Similar to $M_5=1$ for the free case of the DWF, $4-3u_0$ is the
critical value for $M_5$ when coupled to the gauge fields. In order to suppress the effects of this
unphysical mode, its absolute energy $\hat{E}$ should be lifted up high enough so as to be effectively decoupled from
the physical particles. This can be realized by choosing $M_5$ in the vicinity of $4-3u_0$. For
$u_0\sim 0.85$, the critical $M_5$ is approximately $M_5^{\rm {(cr.)}}\sim 1.45$.
On the other hand, in order for the domain wall fermions which are well localized on the domain wall and antidomain
wall such that the explicit breaking of chiral symmetry owing to the finite extension of the fifth
dimension is small, one requires that $b(p)$ be close to zero for the case of $p_\mu \sim 0$.
This gives
\begin{eqnarray}\label{critical}
b(p)&=&1-M_5+\sum\limits_\mu (1-u_0\cos(p_\mu))\nonumber\\
&=&5-M_5-4u_0+2u_0\sum\limits_\mu\sin^2\frac{p_\mu}{2}\sim 0.
\end{eqnarray}
Thus, for $u_0\sim 0.85$, $M_5\gtrsim 1.6$ is preferred.  A chiral window around $M_5\sim 1.5$ is also observed
in the mean-field analysis of the effective mass of chiral modes~\cite{Feng06, Aoki:1999} .
This coincides with the observation in the tuning of $M_5$ initiated by the Columbia
group~\cite{Christ:1998}. The tuning pays attention to the quark condensate $\langle
\bar{\psi}\psi\rangle$ when varying $M_5$ and finds a window of $1.65<M_5<2.15$ where $\langle
\bar{\psi}\psi\rangle$ is nonzero and insensitive to $M_5$. In the mean time, they also observed
the presence of a translational invariant mode in the fifth dimension which contributes about 1\% to
the value of chiral condensate when $L_s=90$. This, we believe, is surely contributed by the unphysical
mode we
discussed above. Since the constant mode is vector like and is normalized by $\sqrt{L_s}$, their
contribution to the quark condensate is suppressed by $1/L_s\sim 0.01$ for $L_s=90$.

The other meaning of this critical value,$M_5^{\rm (cr)}$, is that the oscillatory unphyiscal mode
will appear when $M_5$ is larger than it. Practically, $M_5$ is tuned case by case for different
gauge actions (and different lattice spacings). In the realistic lattice QCD simulations, $u_0$ is
usually in the range of $0.8-0.9$. For example, the $u_0$ of the RBC\&UKQCD $24^3\times 64$
ensemble~\cite{RBC_review2010} with $m_s=0.04$ is roughly 0.85 and their $M_5$ takes the value, $M_5=1.8$; thus,one has
the energy of this mode:
\begin{equation}
\hat{E}=\ln\left|\frac{4-M_5-3u_0}{u_0}\right|\sim 0.89.
\end{equation}
Note that this mode is independent of the bare quark mass. In order for these modes to be much
higher than the physical modes, one has to tune the parameter $M_5$ with the given gauge action and
the bare coupling constant $\beta$.

\subsection{Domain wall valence quarks}

There are many lattice studies using the domain wall fermion operator for valence quarks in the quenched approximation or in mixed action formalisms.
In these cases, for a given gauge
configuration, the valence quark propagator in the real spacetime can be written as
\begin{equation}\label{full-prop}
S_F(\mathbf{x},t;\mathbf{0},0)=S_F^{(\rm ph)}(\mathbf{x},t;\mathbf{0},0)+S_F^{(\rm
un)}(\mathbf{x},t;\mathbf{0},0),
\end{equation}
where $S_F^{(\rm ph)}$ is the contribution from the physical mode, and $S_F^{(\rm un)}$ accounts
for the contribution from the unphysical mode.

If some source technique is applied so that the operator mostly couples to the physical ground state, one has
\begin{equation}
\label{oscillation}
 C(t)\approx W_1 e^{-m_gt}+ W_2(-1)^te^{-Et},
\end{equation}
where $m_g$ is the ground state mass. After some simple calculus,  the effective mass function can be written as (if we consider ${W_2}/{W_1}$ to be somehow small in this situation)
\begin{equation}
M_{\rm eff}(t)\equiv \ln \frac{C(t)}{C(t+1)} \approx m_g+\frac{W_2}{W_1}(1+e^{-\Delta})(-1)^t e^{-\Delta t},
\end{equation}
where $\Delta \equiv E-m_g$. This is exactly the oscillatory behavior in the effective mass plots: i)
the factor $(-)^t$ gives the oscillatory behavior; ii) the amplitude of the oscillation is dictated
by the ratio $W_2/W_1$; and iii) $\Delta$ is the exponential damping parameter of the oscillatory behavior.

\subsection{Domain wall sea quarks}
In the path integral formalism, it is known that the sea quarks enter the expectation value of any physical observable through the fermion determinant, which includes all the possible sea quark loops. This is the reason why a hadronic two-point function can have the contribution from multihadron states.  With the presence of the DWF sea, there exist both physical and unphysical quarks in the sea. Even with a quark bilinear operator $O^{\rm{(ov)}}$ constructed by the overlap
fermion fields, intuitively one expects the following Fock states when coupled to the vacuum
\begin{eqnarray}
O^{\rm{(ov)}}|0\rangle &=& c_1 |\bar{q}_v q_v\rangle + c_2 |\bar{q}_v q_v
\bar{q}_s^{\rm{(ph)}}q_s^{\rm{(ph)}}\rangle \nonumber\\
&+& c_3 |\bar{q}_v q_v \bar{q}_s^{\rm{(ph)}}q_s^{\rm{(un)}}\rangle+\ldots,
\end{eqnarray}
where $q_v$ denotes the valence quark, $q_s^{\rm{(ph)}}$ the physical sea quarks, and $q_s^{\rm{(un)}}$ the
unphysical sea quarks. The third state would give the oscillatory behavior of the correlation function, which is similar to Eq.~(\ref{oscillation}), so the final effective mass function we get will surely contain an oscillatory term.

As is addressed above, with the working value of $M_5\sim 1.8$, unphysical modes exist both for
valence and sea domain wall fermions, which contribute to the hadronic two-point functions by
oscillatory terms in time. This kind of unphysical mode has complex effective energies of
$O(1/a)$. For fine lattices with small $a$s, these modes are heavy and the spurious states involving
them lie much higher than the lowest-lying few states. However, for coarse lattices, these modes
are not that heavy. Taking the vector meson of the light quarks, for example, its mass is roughly 0.77
GeV, and the first radial excitation has a mass of roughly 1.5 GeV. If the unphysical quark does
not make a bound state with the physical quarks, the spurious state is expected to have a mass
approximate to the effective energy of the unphysical quark, $E\sim 1/a$. So in the case of
$1/a\sim 1.5$ GeV, the spurious state can be in the vicinity of the physical excited states.
In other words, when domain wall fermions are involved, one must be
cautious in the data analysis
of two-point functions when considering the excited states, especially for coarse lattices.

\section{Conclusion}\label{section4}
In this work, we explore the origin of the oscillatory behavior of hadronic correlation functions
observed in the practical lattice study when generic domain wall fermions are involved. Even though it is well
known that the zero modes of the four-dimensional Wilson fermion operator $D_W(-M_5)$ with a negative
mass $-M_5$ correspond to nondamping (constant) propagations along the fifth dimension $s$, this
is not the whole story. In the lattice formulation of the free domain wall fermions, it is found that
the zeros of the $D_W(-M_5)$ in the four momentum space give rise to singularities of the DWF propagators
in the $s$-dimension, which is also the origin of the unphysical massive modes in the physical spacetime and
is described by the dispersion relation $1+b^2(p)+\tilde{p}^2=2b(p)$.
In the static case $p_i=0$, the effective energy of this kind of mode is $E=\ln(1-M_5)$. Obviously
for $M_5>1$, $E$ acquires an imaginary part $i\pi$ which accounts for the oscillatory behavior of the
fermion propagator in the temporal direction. For the case with gauge fields present, we adopt the mean-field
approximation and get the static effective energy
$E=\ln[(4-3u_0-M_5)/u_0]$, where $u_0$ is the quantum average of gauge links and takes a value $0.8\sim 0.9$
for the parameters in the typical practical lattice simulations. In this case, the critical value is $M_5\sim 1.5$, above
which the oscillatory behavior appears.

In the practical lattice simulation with the DWF action, $M_5$ is usually tuned by requiring a small
explicit chiral symmetry breaking signaled by the residual mass. This demands the chiral modes bound on the domain walls to damp as soon as possible
along the $s$-dimension. In the mean-field approximation, it is found that this requirement can be reached if the working
$M_5$ is taken to be close to the critical value $5-4u_0$ from Eq.~(\ref{critical}). For the typical values of $u_0$, this
critical $M_5$ is roughly 1.6. This explains to some extent the reason why the working $M_5$ is set in the range $1.7\sim 1.9$
in the realistic studies.

We also discuss the possible consequence of the unphysical fermion modes in the real study when domain wall fermions
are involved either as the valence or sea quarks or both. They do contribute to the hadronic correlation functions
as oscillatory terms with the working $M_5$ parameter. Since their effective energy, $E$, is of $O(1)$ in the unit
of the lattice spacing, the corresponding unphysical states do not lie high enough to avoid entangling
with the excited physical states. This poses a challenge to discern the physical states from the contamination of the
unphysical modes in the study of the hadron spectrum.

\section*{ACKNOWLEDGEMENTS}

This work is supported in part by the National Science Foundation of China (NSFC) under Grants
No. 10835002, No. 11075167, No. 11105153, and No. 11335001 and also by the U.S. Department of Energy Award No. DE-FG05-84ER40154.
Y.C. and Z.L. also acknowledge the support of NSFC and
DFG through funds provided to the Sino-German CRC 110, ``Symmetries and the Emergence of Structure
in QCD."


\end{document}